# Stochastic modeling of scientific impact

M.V. Simkin

Department of Electrical and Computer Engineering, University of California, Los Angeles, CA 90095-1594

Recent research has found that select scientists have a disproportional share of highly cited papers. Researchers reasoned that this could not have happened if success in science was random and introduced a hidden parameter Q, or talent, to explain this finding. So, the talented high-Q scientists have many high impact papers. Here I show that an upgrade of an old random citation copying model could also explain this finding. In the new model the probability of citation copying is not the same for all papers but is proportional to the logarithm of the total number of citations to all papers of its author. Numerical simulations of the model give results similar to the empirical findings of the Q-factor article.

## Introduction

Sinatra et al (1) analyzed a large set of citation data. They observed that some scientists have a disproportional share of highly cited papers. They reasoned that this could not have happened if success in science was fully random. To explain this, they introduced what they called the Q-model. It states that the impact of a paper (they use the number of citations a paper gets during first 10 years since publication and denote it $c_{10}$ ) is given by

$$c_{10} = QP \qquad (1)$$

Here *P* is a lognormally distributed random variable and *Q* is fixed for each scientist but lognormally distributed among different scientists. The authors with high *Q* have many highly cited papers. The *P*-factor describes the role of randomness or luck and *Q*-factor describes talent or individual ability. Sinatra et al (1) compare their experimental findings that are well fitted by the Q-model with what they call a Random-impact model (or R-model). The R-model is like Eq. (1) without a Q-factor. That is impact of a paper is randomly selected out of the impacts of all papers by all scientists. The R-model contradicts experimental data since it does not allow select scientists to consistently produce high-impact papers.

Regarding their R-model Sinatra et al (1) do not refer to any article that argued that success in science could be random. So, it is not clear from the paper whom they argue with. Bibliometrics scholars used stochastic models for long time (2). Randomness was implicit in such models. Silagadze (3) even explicitly titled a chapter of his paper dealing with a random citation model "The almighty chance." Stochastic approach to citations gained firmer footing after statistical studies of misprints propagation in references had uncovered that about 80% of scientific citations are copied from the lists of references used in other papers (4,5). This finding stimulated the model of random citing scientists (6) which is as follows. When a scientist writes an article, he picks up several random papers, cites them, and also copies a fraction of their references. The model leads to cumulative advantage (7) or preferential attachment (8) process, so that the rate of citing a particular paper is proportional to the number of citations it has already received. The model accounted quantitatively for the empirically observed highly skewed distribution of citations. It could explain why some papers got cited thousand times more than the others using the simple law of chances rather than talent. So the success in science could be completely random.

However empirical findings of (1) contradict the predictions of the model of random-citing scientists (6). It does not have a mechanism letting select scientists get many highly cited papers. Its predictions are the same as those of the R-model of (1). In present paper I put forward an upgraded model where the probability of citation copying is not the same for all papers but is proportional to the logarithm of the total number of citations to all papers of its author. Numerical simulations of the model give results similar to the empirical findings of (1).

## Selective Citation Copying Model

Before defining the model that I will be using in this letter it could help to review its precursors. In the original model of random citing scientists (6), when a scientist writes an article, he picks up several random papers, cites them, and also copies a fraction of their references. The model accounted for empirically observed power-law distribution of citations. We should mention that some papers ((1), for example) state that the citation distribution is lognormal. Note, however, that in log-log coordinates power-law distribution is a straight line, while lognormal is a parabola. So lognormal distribution will always give a better fit due to an extra parameter. The difference between the two distributions is not important here. What is important is that both distributions are highly skewed and produce occasional papers with thousands of citations while most papers have only few of those if any.

The model of random-citing scientists had a defect for it produced an exponential distribution of citations to the papers of the same age. It gave an overall power-law distribution of citations because within its framework the highly cited papers were the old ones. In practice the distribution of citations to the papers of the same age is highly skewed and well fitted by a power-law or a lognormal distribution. The distribution of $c_{10}$ of (1) is an example of the distribution of citations to the papers of the same age.

The modified model of random-citing scientists (9) fixed the above problem. It is as follows. When a scientist writes an article, he picks up several random *recent* papers, cites them, and copies a fraction of their references. The difference with the original model is the word "recent." This minor verbal change led to a major difference in mathematical behavior and the model produced the required distribution of citations to the papers of the same age. Another accomplishment of the model was the explanation of the fact that articles get most of their citations in few years following their publication. It was a natural consequence of copying from recent papers. Cumulative advantage model on its own cannot account for that and one must add to the equation an artificial exponentially decaying parameter to match the experimental data (10).

Another addition to the model made in (9) was the fitness parameter for the papers. It was needed to account for certain features of citations distribution I will not dwell upon here. The probability of copying a citation now was not the same for all papers, but proportional to the fitness parameter which was different for different papers.

Here I will use a new definition and interpretation of that fitness parameter. I take it proportional to scientific respectability of paper's author (I ignore the vagaries caused by co-authorship in this study). I will take this scientific respectability equal to the logarithm of the total number of citations, $\log N_k$, the scientist, $k$, had received at the moment. In contrast with (9) now fitness is a dynamic variable since $N_k$ increases with time. It is also not paper-specific, as it was in (9), but author-specific. So, all papers by the same author get the same fitness. The model resembles preferential attachment models with a nonlinear kernel (11).

Variants of the model are possible. $N_k$ may be not the total number of citations but the number of citations to papers published during recent years. Functions other than logarithm are also possible. The logarithm makes sense since it considers that the change in the number of citations from, for example, 100 to 1000 produces a bigger change in scientific respectability than the change from 1000 to 1900. Another justification for the use of a logarithm of the citation number is that it was used by Ioannidis et al (12) when computing their composite citation indicator. With little verbal modifications one can apply the model to writers, musicians, and other cases where select authors have a disproportional share of popular products (13). Instead of citations one can talk about critic reviews.

## Numerical Simulation of the Model

I did a series of simulations of the Selective Citation Copying Model. The model is an extension of the modified model of random-citing scientists (9). It becomes more tractable if we discretize time as it was already done in (9). As a unit we will take a typical time required to publish an article. At the step zero $N_{sci}$ scientists each publish $N_{pap}$ papers. At step one they produce $N_{cit} = N_{sci} \times N_{pap}$ citations to those papers fully at random. We compute the total citation number, $N_k$, for each scientist $k$. We also compute $N_{max}$, the maximum $N_k$. At the step two we select a step one citation at random, look up the $N_k$ of its author, and with probability $\log N_k / \log N_{max}$ copy it to step two. With the remaining probability we select another step one citation at random and repeat the previous step. We keep repeating this until one citation is finally copied. We repeat the procedure $N_{cit}$ times to get $N_{cit}$ step two citations. We re-compute $N_k$ for each scientist and proceed to step three. We repeat the procedure for $N_{st}$ steps. A reasonable approximation for the time required to publish an article is 6 months. So, if we run the simulation for $N_{st} = 20$ we get $c_{10}$.

The above simple model already lets select scientists get many high-impact papers. However, there are some discrepancies with experimental results. For instance, the plot analogous to that of Figure 2(a) is asymmetric. A small modification fixed this problem. 90% of citations are copied exactly as described before. But 10% of the citations are self-citations. That is, they go to randomly selected papers of the same scientist.

Now I shall present the results of a typical outcome of a simulation of a sample of 1,000 scientists each of whom published 25 papers. First I shall look at 25 top cited papers which is the top 0.1% of the sample. It turns out that one scientist has 2 papers on the shortlist, and one got even 3. This is not a likely outcome for the R-model: using standard formulas for a Binomial distribution one gets that the probability to get one scientist with three papers on the shortlist is only 0.2%. In contrast, in simulations of Selective Citation Copying Model I am getting at least one scientist with at least 3 papers on the shortlist in over half of the simulations. Sinatra et al (1) did not use this simple metrics so to compare my results with theirs I have to use their tools.

For every scientist in their sample Sinatra et al (1) found the paper with highest ten-year impact factor $c_{10}$ and denoted it $c_{10}^*$. Afterward they averaged $\log c_{10}$ for the rest of scientist's papers and denoted it $\langle \log c_{10}^- \rangle$. They plotted $\log c_{10}^*$ versus $\langle \log c_{10}^- \rangle$ in Figure 3(D) of (1). I reproduce their results, extracted using Web Plot Digitizer (14), alongside mine in Figure 1(a). Sinatra et al (1) compared their data with the predictions of R and Q models. I reproduce their analysis using my simulation data in Figure 1(b). Note that Sinatra et al (1) incorrectly computed[1] the R-model curve in their Figure 3(D). In Figure 1(b) I plot the R-model curve computed both the right and wrong ways to allow the comparison with Figure 3(D) of (1). As one can see, Q-model fits the data better than the R-model though the difference between the two is much less when you compute the R-curve the right way.

---

[1] The blue curve (R-model) in Figure 3 (D) of (1) is wrong. In the Q-model a higher citation rate for the bulk of the work implies a higher Q and thus a higher citation rate of the highest cited paper compare to the R-model. Thus, the red curve (Q-model) should be above the blue curve (R-model) at the high average impact end. By a similar argument, the red curve should be below the blue curve at the low average impact end. When we look how they compute the R-model curve in Eq.(S9) we see that they average citations for the bulk of the work for fixed number of citations to the highest cited paper. They should have fixed the number of citations per paper for the bulk of the work and computed the average number of citations to the highest cited paper. The "Wrong curve" in Figure 1(b) was computed using the wrong method of (1) using my simulation data.

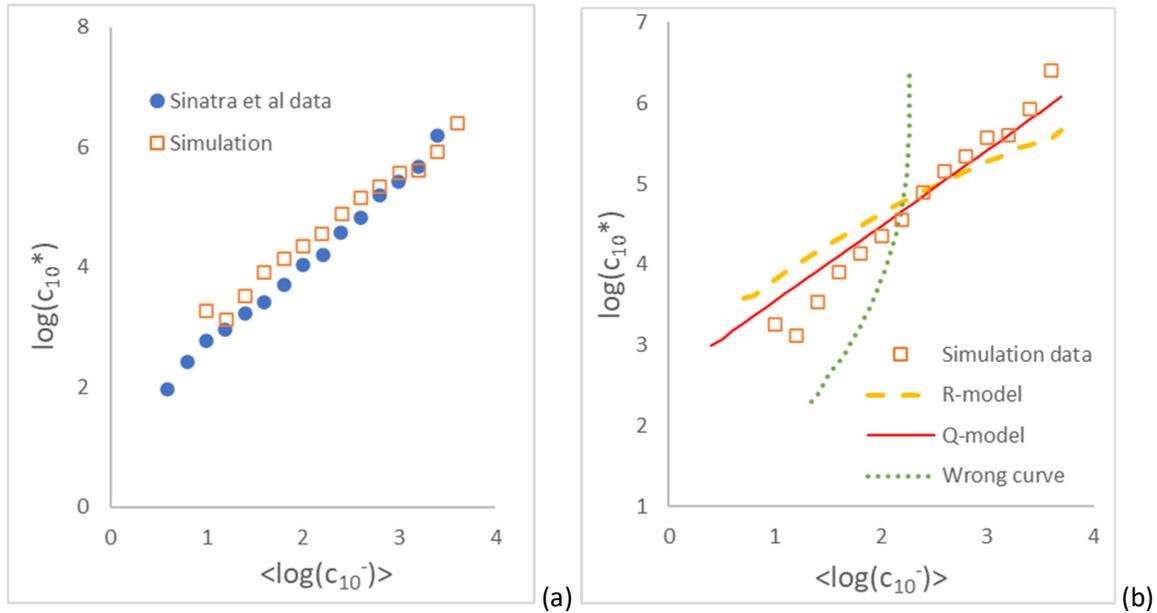

**Figure 1.** Highest impact paper versus the average impact of the bulk of scientist's work. (a) Comparison of my simulation results with the data of Figure 3(D) of (1). (b) Comparison of the simulation results with the predictions of R and Q models.

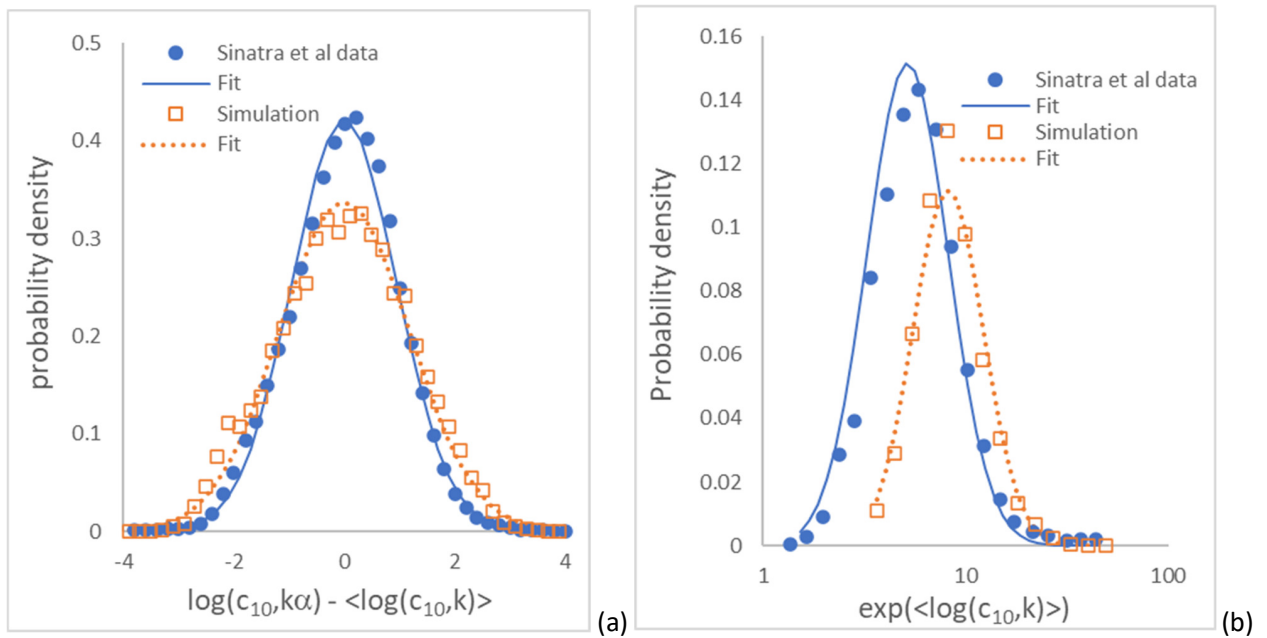

**Figure 2.** (a) Distribution of the deviations of of the impact of individual scientist's papers from scientist's impact geometric mean. Comparison of the simulation data with the data of Figure 3(F) of (1). The fitting curve is a normal distribution with mean and variance equal to those of the simulation data. (b) Distribution of the geometric means of scientist's papers impacts. Simulation data compared with the data of Figure 3(G) of (1). The fitting curve is a lognormal distribution with mean and variance equal to those of the simulation data.

I also computed approximate P and Q distributions for the simulation data. Following (1) I computed for each scientist $k$ the logarithm of the geometric mean of the impacts of their papers, $\langle \log(c_{10}, k) \rangle$. And plotted the distribution of the deviations of the impacts of the individual papers $\log(c_{10}, k\alpha) - \langle \log(c_{10}, k) \rangle$ in Figure 2(a). Here $k\alpha$ denotes paper $\alpha$ of scientist $k$. The fitting curve shown in the plot is a normal distribution with mean and variance equal to those of the simulation data. For comparison I include the corresponding data extracred from Figure 3(G) of (1).

I computed for each scientist the geometric mean of their papers impacts. I plot the distribution of these means in Figure 2(b). For comparison I include the corresponding data extracted from Figure 3(G) of (1). To be precise I also used Eq.(3) of (1) to rescale the extracted data.

## Discussion

The model that uses no talent parameter can produce high-impact scientists. It also reproduces qualitatively some intricate properties of the actual citation distribution. Could it describe what is actually going on? This suggestion may meet with rash criticism so I will supply some corroborating evidence that gives it some credence.

The original push for the model of random-citing scientists was the discovery of copied citations identified through the propagation of identical misprints (4,5). There is more evidence pointing to the same end.

In a recent development a non-existing paper got 400 citations. Harzing (15) found out it was a made-up reference used in a publisher's template for conference proceedings to illustrate reference style. The scientists used the template to write their articles and forgot to delete the sample reference. An interesting observation is that not all of those citations appeared in conference proceedings where the template was used. Apparent reason: citation copying.

Hamilton (16) had studied 47 retracted articles and discovered that 34 of them received over 500 citations after retraction. 92% referenced retracted articles as legitimate work. The apparent reason is again citation copying.

Another argument is that citation numbers do not match well with Nobel Prize winning. Ioannidis et al (12) compiled a list of 100,000 most cited scientists using all indexed by Scopus articles published between 1997 and 2017. Kosmulski (17) reported that out of 97 recent Nobel prize winners only 90 were on that list. And only 45 were among the top 6,000.

I checked the same data set (12) and discovered that it includes only 6 out of 22 Abel Prize winners. Andrew Wiles who proved Fermat theorem is prominently missing. Only 22 of 60 Fields Medal recipients are on that list. This suggests that if the Q-factor is indeed talent – it surely is not a mathematical one.

Note, however, that 27% of Abel Prize winners enter 0.93% of top cited mathematicians (for that is the percent of mathematicians that made it into the list of 100,000 top cited scientists). This could not have happened had scientific citing been completely random. The estimation of the relative role of scientific contribution and other factors to citation numbers will be the subject of another paper.